# GNSS Time Synchronization in Vehicular Ad-Hoc Networks: Benefits and Feasibility

Khondokar Fida Hasan, *Student Member, IEEE*, Yanming Feng, and Yu-Chu Tian, *Member, IEEE*

*Abstract*—Time synchronization is critical for the operation of distributed systems in networked environments. It is also demanded in vehicular ad-hoc networks (VANETs), which, as a special type of wireless networks, are becoming increasingly important for emerging cooperative intelligent transport systems. Global navigation satellite system (GNSS) is a proven technology to provide precise timing information in many distributed systems. It is well recognized to be the primary means for vehicle positioning and velocity determination in VANETs. However, GNSS-based time synchronization is not well understood for its role in the coordination of various tasks in VANETs. To address this issue, this paper examines the requirements, potential benefits, and feasibility of GNSS time synchronization in VANETs. The availability of GNSS time synchronization is characterized by almost 100% in our experiments in high-rise urban streets, where the availability of GNSS positioning solutions is only 80%. Experiments are also conducted to test the accuracy of time synchronization with 1-PPS signals output from consumer-grade GNSS receivers. They have shown 30-ns synchronization accuracy between two receivers of different models. All these experimental results demonstrate the feasibility of GNSS time synchronization for stringent VANET applications.

## I. Introduction

COOPERATIVE Intelligent Transportation Systems (C-ITS) are an emerging and multidisciplinary area with increasing importance and wide applications. They create an environment of information sharing and activity cooperation for increased road safety and system operation efficiency [1]. A key component of a C-ITS is Vehicular Ad-hoc NETwork (VANET). Naturally, a VANET consists of a large number of network nodes that pass each other at a high speed. Thus, it is a highly dynamic and decentralized network. It shows unique characteristics in comparison with other types of wireless networks. A C-ITS crucially relies on VANETs for real-time wireless communications of various data generated from in-vehicle and roadside sensors. Therefore, the functionality and performance of VANETs determine how well a C-ITS can facilitate various applications on road [2]–[4].

Manuscript received June 10, 2017; revised November 10, 2017; accepted December 26, 2017. This work was supported in part by the ICT-Division, Bangladesh, and in part by the Australian Research Council under Grant DP160102571 and Grant DP170103305. The Associate Editor for this paper was H. A. Rakha. *(Corresponding author: Yu-Chu Tian.)*

The authors are with the School of Electrical Engineering and Computer Science, Queensland University of Technology, Brisbane, QLD 4001, Australia (e-mail: y.tian@qut.edu.au).

Color versions of one or more of the figures in this paper are available online at http://ieeexplore.ieee.org.

Digital Object Identifier 10.1109/TITS.2017.2789291

In wireless communications, time synchronization is essential for coordination and consistency of various network events. It is also necessary for accurate sequencing and real-time control tasks. In general, all physical clocks run at different rates from a reference clock, and drift apart from other clocks. Time synchronization is a technique to keep clocks remain accurate. In a network of nodes, it adjusts clock drifts, and synchronizes all clocks with a globally known time or with each other. This ensures every node in the network to operate with the same notion of time.

While VANET communications are considered to be asynchronous in nature, time synchronization among vehicles is essential for many applications [5]. This is similar to the Internet, which is embedded with time synchronization mechanisms. Some VANET applications are highly time-sensitive. In these applications, maintaining a real time and also a precise timing between communicating nodes is critical. Therefore, as in other synchronous wireless distributed systems, many VANET applications depend on synchronous communications to provide common C-ITS services. Examples include coordination of activities [6]–[8], relative vehicle positioning, data communications, and security services [9]. Moreover, in order to record event information over the network, a VANET needs to maintain an accurate physical time. This also demands maintenance of an accurate standard time through time synchronization in an asynchronous manner.

Modern vehicles are already integrated with a GPS (Global Positioning System) or multi-GNSS (Global Navigation Satellite System) receiver for navigation. Accurate time support from GNSS is therefore plausibly achievable. While GPS-based time synchronization is used in many networks, limited reports have been found on test results of time synchronization from GPS or GNSS receivers. Despite the modernization of GPS receivers, it is still not well understood whether or not GNSS time synchronization is beneficial and feasible. Can consumer-grade GNSS receivers provide the time synchronization availability and accuracy required for VANETs? In a VANET environment, is GNSS time synchronization better than existing decentralized time synchronization methods originally developed for other wireless networks?

This paper aims to answer these questions. It begins with a discussion of the significance and benefits of time synchronization in VANETs. Then, it compares existing time synchronization techniques developed for general mobile ad-hoc networks (MANETs). After that, it reviews existing efforts for VANET time synchronization and highlights the feasibility of GNSS-based time synchronization for VANETs. Experiments are conducted on coverage availability and timing accuracy of GPS time solutions to support our claim on the feasibility of GNSS-based VANET time synchronization.

## II. Significance and Requirements of Time Synchronization

Time synchronization is one of the key issues for successful operation of communication networks. It enables network nodes to maintain an accurate and precise time. As a result, accurate time-stamping and meaningful ordering of transmitting messages become possible. This allows seamless network operation and Quality of Service (QoS) of various network events in a practical communication network system. Therefore, time synchronization improves network inter-operability and meaningful coordination among network devices, and consequently enforces end-to-end connectivity of networks.

One of the expectations from time synchronization is to eliminate the latency in critical data transmission. This is achieved through scheduling a traffic lane for transmitting packets. The traffic lane is a commonplace in a wireless network, regardless of a synchronous or asynchronous transmissions. Along with channel scheduling, various network applications also rely on time synchronization. In some network applications, network nodes only need relative time synchronization for ordering various network events [10]. In this case, the clocks of the nodes are synchronized with each other irrespective of the high accuracy of the synchronization. In many other network applications, all network nodes need their clocks to be synchronized with a highly accurate clock.

VANET is a soft real-time communication technology with strict time boundaries for end-to-end transmission delay. Some applications characterize the time-sensitive nature of VANET. First of all, VANET is very dynamic with vehicles coming in and moving out, making relative time synchronization more difficult in VANET than in other general networks. Secondly, for vehicle safety applications, vehicle location and velocity data, as part of Basic Safety Messages (BSM), need to be exchanged frequently, e.g., in 10–100 Hz, between vehicles over a single hope or multiple hops. Furthermore, event-driven safety messages to be transferred over VANET are highly time-sensitive. For example, Wireless Access for Vehicular Environment (WAVE) short messages about accidents, stop/slow vehicle warning, Blind Spot Warning (BSW), and Emergency Electronic Brake Light (EEBL) are required to be broadcast to targeted nodes within a fraction of a second. A typical end-to-end network latency is up to 100 ms for many VANET applications [11]. Any further offset will expose vehicles in a risky environment. Thus, precise clocks are expected in VANET, demanding accurate time synchronization.

So far, there has been limited work [12], [13] on time synchronization and stringent timing requirements for VANET. The concepts and techniques of time, time quality and time synchronization in VANET have been directly adopted from Wireless Local Area Network (WLAN) standards. However, WLAN is an infrastructure centered asynchronous network, in which communications are implemented among low mobility wireless nodes through fixed Access Points (APs). As a WLAN is not time critical, precise time synchronization is not required nor is really achieved. In comparison with WLAN, VANET consists of both infrastructure-based vehicle-to-infrastructure (V2I) and ad-hoc-based vehicle-to-vehicle (V2V) communications. The integration of V2I and V2V in make VANET more challenging and time critical than WLAN.

The benefits and requirements of time synchronization in VANET are further recognized from the following two categories. The first category is VANET applications in which time synchronization is essential. The other category is VANET applications in which time synchronization is desirable, which enhances system performance or resource utilization. These two categories are discussed below in detail.

### A. Applications Where Time Synchronization is Essential

*1) Network Interoperability and Coordination:* indicate the capability of the networks to send and receive messages and communicating information among the inter-connected networks, devices and nodes. It is the ability of efficient and meaningful coordination among networking nodes and components for information exchange.

*2) Scheduling of Channels:* is required for efficient use of channel resources. VANET utilizes short range communications, e.g., 5.9 GHz Dedicated Short-Range Communication (DSRC) technology, typically within a range of 1 km to provide high data rate and low latency. In general, a trade-off exists between the efficiency and reliability of VANET communications. The efficiency is typically characterized by bandwidth consumption, channel utilization, and channel coordination. Therefore, designing an efficient WAVE Medium Access Control (MAC) protocol is essential for improved efficiency, enhanced QoS, and reliable packet transmission. Time synchronization plays a crucial role in MAC coordination. Existing IEEE 802.11p MAC uses Time Synchronization Function (TSF) to coordinate channels. The underlying medium access mechanism is Carrier Sense Multiple Access/Collision Avoidance (CSMA/CA), which is asynchronous in nature [14] and is only applicable to those systems in which precise sub-second timing is not required. In dense VANET scenarios, CSMA/CA does not support highly accurate time synchronization.

*3) Road Safety:* is a critical objective in VANET. To avoid unpredictable events that may cause road accidents and causality, VANET network delay is expected to be small and predictable. Effective channel scheduling will help reduce the unpredictability of VANET communications. In compared with CSMA/CA, time slotted access protocols offer collision-free communications with predictable dynamics. Slotted protocols, e.g., STDMA, MS ALOHA, RR ALOHA and UTRA TDD, show good scalability, high reliability, and fair use of channel resources [8], [15]–[21]. However, these time slotted protocols require precise absolute time synchronization for time slot coordination [22].

*4) Security:* is a significant concern in VANET. Session hijacking and jamming are two communication threats for the manufacturer, forensic experts and transport regulation authorities. Precise time synchronization is a key tool for development of traceable and reliable communications. This allows reconstruction of the packet sequence on the channel, and thus effectively helps overcome the threats. It is indicated that a fine-grained analysis of channel activity between concurrent transmissions requires stringent timing guarantees of $8\mu s$ for DSRC [23]. A well-known approach to prevent wormhole attack is packet leash, which also requires highly accurate clock synchronization [24].

*5) GPS-Based Relative Vehicle Positioning:* requires time synchronization. Time-to-Collision (TTC) on roads depends on relative locations of two vehicles. The most stringent requirements for relative positioning accuracy is about 10 cm [25]. Vehicle state and time information are exchanged between vehicles to compute relative vehicle positions for safety decisions. For a vehicle travelling at the speed of 110 km per hour, a timing error of 10 ms will cause a position uncertainty of 30 cm, which is too high for collision avoidance. To meet the accuracy requirement of 10 cm, it is required to keep timing errors within 3 ms, giving a relative positioning error of about 9 cm. This requirement does not look very high. However, it emphasizes that time synchronization is essential.

### B. Applications Where Time Synchronization is Desirable

*1) A Guard Interval:* is used between two time slots in slotted access protocols. Packet propagation starts at the beginning of a new slot after a guard interval. This helps accommodate timing inaccuracy and propagation delay. The guard interval should be set to be bigger than the worst time synchronization accuracy for the slotted access mechanism to work properly. Precise absolute time synchronization helps reduce the guard interval greatly, thus increasing the channel slot duration significantly [26]. For example, the time slotted access protocol STDMA is used in ship navigation system through Automatic Identification System (AIS). A typical frame length of STDMA in AIS is 2,016 slots. It is shown [27] that compressing the guard interval by 10 $\mu s$ will accommodate 45 new slots of 496 $\mu s$ each. This is translated to a noticeable increase in channel capacity, implying that more time slots can be used for packet delivery.

*2) Localization:* is a typical application in VANET. It is direcctly related to the challenging issue of determination of accurate ranges and range rates of vehicles to other vehicles. Terrestrial radio frequency based ranging techniques, e.g., Time of Arrival (TOA) and Time Difference of Arrival (TDOA), are effective ways for location determination [28]–[30]. They essentially involve distance calculation involving the speed of light. This implies that a timing accuracy of 10 ns corresponds to a distance measurement accuracy of about 3 m. If time synchronization is accurate to 1 ns, the ranging accuracy can be improved to 0.3 m. Therefore, highly accurate absolute time synchronization will enable to use terrestrial radio ranging signals in vehicle location determination.

### III. Time Synchronization Techniques in MANET

Mobile Ad hoc Network (MANET) is formed by a cluster of spontaneously communicating wireless mobile nodes without any pre-designed network infrastructure. This unique characteristic indicates that the bandwidth, power and network configuration are important factors in its operation. Time is a critical factor for coordination of all network events and resources. Therefore, maintaining an accurate time among all network nodes is a significant issue in MANET [31].

Time synchronization techniques used in traditional wireless networks are not appropriate for MANET [32]. In traditional networks, message delay can be estimated with a high degree of accuracy. However, in MANET, an intermediate node may cause a long delay in the overall message transfer. Furthermore, periodic broadcasting of messages is not realistic in MANET due to limited energy and network resources.

Efforts have been made in time synchronization of dispersed nodes in MANET. Most of these efforts deal with decentralized methods, especially for wireless sensor networks, for relative time synchronization, which takes place with local time via remote inter-process message exchange. In such relative synchronization methods, it is not required to strictly maintain a standard notion (value) of time or to follow any authorized and known global time. Instead, all communicating nodes are synchronized relatively with each other.

Broadly speaking, all such protocols implement either sender-receiver based or receiver-receiver based time synchronization. In sender-receiver based synchronization, a handshaking is established between the sender and receiver. Then, a round-trip delay is calculated [33], [34]. An example of sender-receiver based synchronization is the Timing-sync Protocol for Sensor Networks (TPSN), which provides a time accuracy of 20 $\mu s$ on average [35]. In receiver-receiver based synchronization, a set of receivers are synchronized with one another. In this case, receiver nodes compare the time-stamps of the messages received from the same sender (another receiver). A typical example is Reference Broadcast Synchronization (RBS), a popular synchronization method in wireless sensor networks. RBS behaves with an accuracy of 6.29 $\mu s$ in an average case. This accuracy is primarily determined by the amount of time to receive and process packets. A larger difference in the packet receipt delay will deteriorate the time accuracy [36]. Using a unique combination of the above two methods, Flooding Time Synchronization Protocol (FTSP) is able to provide a time accuracy of 1.5 $\mu s$ on average. For general sensor network applications, such a time accuracy through relative synchronization is sufficient.

Comparisons of various time synchronization protocols in MANET are tabulated in Table I. The average time synchronization accuracy and worst accuracy are also given in the table for each of these protocols.

### IV. Time Synchronization in VANET

While there exist time synchronization protocols for MANET, they are not directly applicable to VANET. This is result from some fundamental difficulties. First of all, VANET nodes are highly dynamic and the relative speed up to 200 km/h. Frequent changes in network topology and links make relative time synchronization very difficult. Also, some vehicle safety applications on road demand extremely small delays in message transfer [42]. This demands highly accurate timing for coordination of network events. Furthermore, existing synchronization protocols are mostly based on a tree topology, but it is not feasible to build a stable tree for synchronization in highly dynamic VANET.

TABLE I
Accuracy (Errors) of Timing Sync Protocols in MANETs

| Protocol | Avg. | Worst |
|---|---|---|
| **Sender to Receiver** | | |
| Timing-sync Protocol for Sensor Networks (TPSN) [35] | 20 $\mu s$ | 50 $\mu s$ |
| Secure Group Synchronization (SGS) [37] | 47.13 $\mu s$ | 130 $\mu s$ |
| **Receiver to Receiver** | | |
| Reference Broadcast Synchronization (RBS) [36] | 6.29 $\mu s$ | 20.53 $\mu s$ |
| Time Diffusion Synchronization Protocol (TDP) [38] | 100 $\mu s$ | unknown |
| Gradient Time Synchronization Protocol (GTSP) [39] | 4 $\mu s$ | 14 $\mu s$ |
| **Hybrid** | | |
| Flooding Time Synchronization Protocol (FTSP) [40] | 1.5 $\mu s$ | 3 $\mu s$ |
| Light Weight Time Synchronization (LTS) [41] | 0.4 s | 0.5 s |

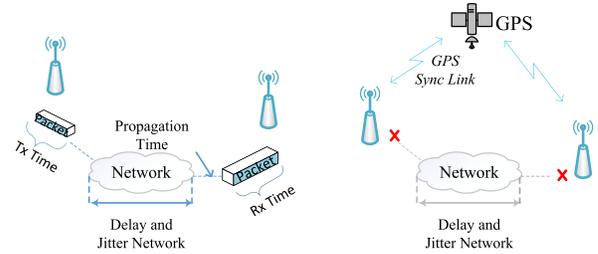

Fig. 1. In-band time synchronization (left diagram), and out-of-band external time synchronization (right diagram).

Overall, time synchronization in VANET has not been well understood. Two main types of approaches, namely centralized and decentralized, give us some hints for potential development of synchronization techniques for VANET.

In a mutual decentralized synchronization method [43], a carrier frequency is continuously devoted to synchronization. The resulting accuracy of synchronization is shown to be within 20 ms. Standard deviation is demonstrated to be below 3% for 99% of the time. Another decentralized method is known as decentralized slot synchronization [26]. Without the need of additional signaling effort, it uses a mutual adaptation of individual slot timing for time synchronization. A further decentralized method is Converging Time Synchronization (CTS) protocol [9]. It does not use any additional physical pulse device, e.g., envelope detector, which is employed in the above two decentralized methods. However, it is superior to other decentralized synchronization protocols for its simplicity and accuracy stability. In CTS, any incoming vehicles get synchronized with the largest synchronized group with a time deviation of 11 $\mu s$.

The necessity of tight time synchronization over VANET is discussed through a comparative analysis between the two air access methods CSMA/CA and MS-Aloha [19], [21]. However, experiments have not been reported to demonstrate the technical analysis. Exploiting GNSS signals, a centralized synchronization solution is proposed for inter-vehicle communications [44]. It uses the framework of the UMTS Terrestrial Radio Access Time Division Duplex (UTRA TDD) as its air interface. Since then, further developments of GNSS-based time synchronization have not been found in the literature.

In the next few sections, we will highlight the feasibility of GNSS-based time synchronization and demonstrate our feasibility analysis through experimental studies.

## V. Feasibility of GNSS Synchronization in VANET

GNSS is a space-based Positioning, Navigation and Timing (PNT) utility alongside wireless wide area networks and communications. It has a potential to bring significant benefits to centralized and accurate time synchronization in VANET.

### A. Justification of the Feasibility

First of all, GNSS services are capable of providing absolute time synchronization support over local and decentralized time synchronization protocols [9]. Such an external synchronization technique is also free from additional message transfer delay between nodes, as shown in the right diagram of Fig. 1. Most time synchronization techniques are designed with message exchanges between nodes. Therefore, they rely on the data communication networks. This is shown in the left diagram of Fig. 1. The performance of in-band internal time synchronization depends on the channel activities, the density of the communicating nodes, and the condition of the networks. However, message delivery in such networks may sufer from a significant latency and jitter [45]–[47]. On the contrary, GNSS based out-of-band external synchronization techniques do not use the communication networks in its operation. It does not use any bandwidth resources of the networks as all nodes are synchronized with external GNSS signals. Thus, its performance is independent of the number of network nodes. It is worth mentioning that in the absence of GNSS signals, in-band time synchronization has to be activated for better time keeping. For instance, Timing Advertisement Frames in DSRC-based VANET may be used to assist in time synchronization if GNSS signals become unavailable.

The feasibility of GNSS time synchronization is also justified by the fact that a single satellite in view is able to provide time solutions. A GNSS receiver normally tracks all satellites in view to obtain pseudo-range and Doppler measurements at each frequency for Position, Velocity and Time (PVT) computing. The time states include clock bias and clock rate. There are basically two modes for estimation of time states: dynamic mode and static mode. The dynamic mode is used in moving platform applications when the position is unknown. In this case, the receiver can compute its own position and time by tracking four or more GPS satellites. The static mode is the preferred mode for applications with a known fixed position. In this case, the receiver can compute time bias and time rate by tracking one or more satellites.

In the dynamic mode with unknown position, the PVT estimation is performed by solving a set of linear observation equations with the least squares approach epoch by epoch. In other words, the 4D states of the vehicle moving platforms are determined without assuming knowledge of the dynamics of the receiver. The performance of the state solutions depends on two factors: the user range equivalent error (URE) for the observation accuracy, and the geometric dilution of precision (GDOP) about the satellite geometry. In general, the PVT solutions obtained under the GDOP $\leq 6$ are considered to be valid and usable. The position and time errors are almost in the same order of magnitude.

In the static mode, if the receiver position is known to a certain accuracy through alternative positioning techniques or predictions, one tracked satellite at a time is still able to provide timing information with a reasonable accuracy [48]. When there are fewer satellites than four in view, or the satellite geometry is very poor, alternative positioning techniques, such as Inertial measurement units (IMU), are normally involved to determine or improve the position and velocity states. As long as there is one satellite in view, the PVT processor can give a time solution. The time solution outage is always much lower than that of the usable PVT solution outages. In fact, most new GNSS receivers are designed to track multiple GNSS signals, i.e. signals from GPS, GLONASS, Galileo and Beidou (BDS) constellations. As a result, multi-GNSS visible satellites would double or quadruple the number of GPS visible satellites. This will reduce the outage of time solutions significantly, thus making GNSS-based time synchronization much more feasible and reliable than before.

The feasibility of GNSS time synchronization is further justified by the fact that absolute time synchronization with an external global time standard is particularly suitable for VANET applications. VANET nodes operate outdoors most of the time. The density of the nodes vary significantly from time to time or from here to there depending on the traffic of vehicles. In-band time synchronization will suffer from a big jitter and variable latency in exchanges of synchronization messages. This difficulty can be easily overcome by using out-of-band external GNSS time synchronization.

Moreover, consumer-grade GNSS receivers are already mounted in most modern vehicles for positioning and navigation. They are ready to provide GNSS timing information for time synchronization without additional hardware cost.

### B. GNSS Timing Information

The timing information provided by GNSS is highly precise and accurate as it is generated from atomic clocks and maintained very stringently. In a GNSS system, there are three scales of time, e.g. GNSS time, satellite time and standard time such as UTC. These times are different from each other [49], [50]. In the satellite time transfer method, the offset between GNSS and UTC time are transmitted to user receivers for correction. As of September 2017, UTC is ahead of GPS bby 18 s whereas TAI is lagged by 19 s as shown in Fig. 2.

However, the satellite system provides UTC time to a ground receiver through necessary adjustment using navigation message as shown in Fig. 3. In general, a typical satellite has an atomic clock of its own time $t^p$. This time is regulated by the earth bound control segment with GPS time $t_{gps}$. Along with $t_{gps}$, the control segment also uploads navigation messages containing $\Delta t_{utc}$, which is regulated by the United States Naval Observatory (USNO). A GPS receiver has its clock system with time $t_r$. With GPS pseudo-range measurements, the receiver can compute clock bias $\Delta t_r$ with respect to GPS time, together with the receiver position states. $\Delta t_r$ is defined as:

$$\Delta t_r = t_r - t_{gps}. \tag{1}$$

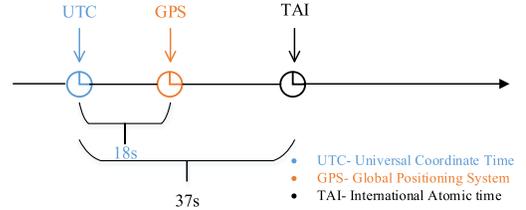

Fig. 2. Time offsets among different atomic scale standards.

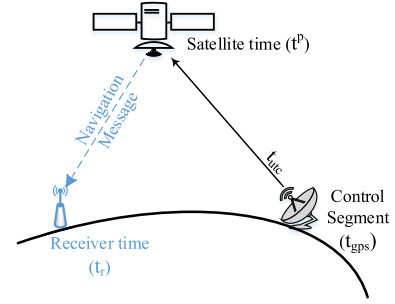

Fig. 3. Time transfer through GNSS [49].

The receiver clock bias in a receiver hardware system is also known as instantaneous receiver clock offset relative to GPS [49]. As the offset can be large, GPS code measurements are known as pseudo-ranges. The clock bias is adjusted to the GPS time when the magnitude reaches a certain limit, such as 0.1s. As a result, a time GPS receiver adjusts the bias and obtains its UTC time using the following relationship:

$$t_{utc} = t_r - \Delta t_r - \Delta t_{utc}, \;\; \Delta t_{utc} = t_{gps} - t_{utc}, \tag{2}$$

where the offset $\Delta t_{utc}$ between GPS time and UTC time can be obtained from navigation messages. $\Delta t_{utc}$ contains the leap seconds (currently 18s) as shown in Fig. 2 and a fractional part since the last leap second adjustment.

With this technique of GPS time transfer, all nodes of a network are individually synchronized with the GPS time. Then, adjust the GPS time with the same offset with respect to the UTC time. As a result, all nodes are synchronized with the UTC time. This is demonstrated in Fig. 4. As shown in Fig. 4 (a) and (b), nodes $N_1$ and $N_2$ are individually synchronized with a satellite ($t_{gps}$), and then are updated to UTC time. Effectively, this will synchronize Node $N_1$ and Node $N_2$ with each other as shown in Fig. 4 (c).

### C. Errors of the Receiver Timing

We now examine the errors of the receiver timing. Only the uncertainty of the GNSS clock bias $\Delta t_r$ will affect time synchronization because $\Delta t_{UTC}$ is common to all VANET nodes. However, the basis for time transfer functions in GPS-based products is one pulse per second (1 PPS) signal supplied by the GPS receiver. This signal is typically a short logic pulse, one edge of which is adjusted by the receiver to be on time with respect to the one second epoch of UTC or GPS time. Errors in the time of occurrence of the 1 PPS pulses from the GPS receiver consist of three parts: 1) bias or offset due to uncompensated propagation and hardware delay errors in the receiver/antenna system; 2) drift, which is the variation in time over an extended period due to changes of satellites tracked over time; and 3) jitter, which is the short-term variation in timing from pulse to pulse. These error sources are inherent in both GPS system and GPS receiver design/implementation. The total effect of these errors is typically tens of nanoseconds to a few microseconds, depending upon the quality of GPS receivers. Receiver manufactures usually calibrate the receiver bias well, yielding a timing accuracy of 10 ns or better under ideal observational conditions. This high level of accuracy is possible because the timekeeping maintained within the GPS system is continuously adjusted to null out timing errors.

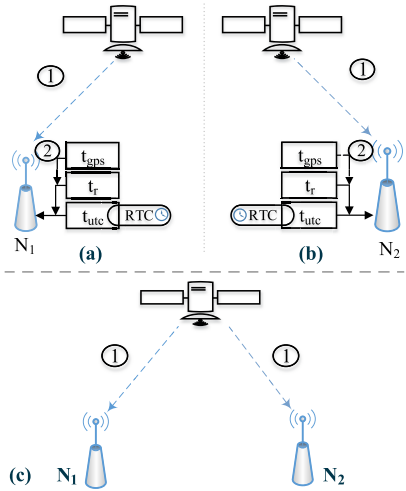

Fig. 4. (a) (b) Nodes $N_1$ and $N_2$ are individually synchronized with GNSS and updated with UTC. (c) Effectively, the two nodes are synchronized with each other via GNSS.

A typical GNSS receiver has an internal quartz based oscillator that continuously runs and follows the GPS time. Currently, a low-end GNSS receiver can update position and time solution at a rate of up to 20 Hz, which corresponds to an interval of 50 ms.

However, such a quartz-based clock deviates over time. This is because the frequencies of the quartz oscillators are different. As time elapses, a quartz clock tends to diverge from the perfect clock, i.e., the real accurate time, and also from each other in a network. Ideally, for a perfect clock, the clock rate of change $dC/dt$ is equal to 1. In practice, this rate may increase or decrease due to the variation of the clock oscillator's frequency. In a quartz clock, this frequency variation is commonplace due to the result of the environmental changes of the node, e.g., variations in temperature, pressure, and power voltage upon the clock. In clock synchronization terminology, this difference in the frequencies of the practical clock and the perfect clock is known as clock skew. The rate of change of the clock skew is known as clock drift. This means that clock drift is the derivative of clock skew [4].

Though the frequency of clock oscillator depends on ambient conditions and may change over time, for a relatively extended period, e.g., minutes to hours, the frequency of node clock can be approximated with good accuracy by an oscillator with a fixed frequency [51], [52]. Therefore, the clock of a node can be expressed as:

$$C_i(t) = d_i \cdot t + b_i, \tag{3}$$

where $t$ is the standard time of the measurement (UTC); $d_i$ is the clock drift due to the oscillator's frequency differences resulting from environmental changes of the node, e.g., variations in temperature, pressure, and power voltage upon the clock; $b_i$ is the initial offset in GNSS synchronization framework, and can be correlated to systematic ranging errors and hardware delays, $d_i$ and $b_i$ can be different from node to node. However, the clock skew $d_i$ is different from the drift in GNSS timing errors offered by GNSS 1PPS outputs. It is also worth mentioning that the derivation of the clock skew can actually be estimated from GNSS Doppler measurements along with the velocity states.

Now, any two such GNSS synchronized clocks (such as Fig. 4(a) and Fig. 4(b)) can be expressed as:

$$\left.\begin{array}{l} C_1(t) = d_1 \cdot t + b_1, \\ C_2(t) = d_2 \cdot t + b_2. \end{array}\right\} \tag{4}$$

By following Fig. 4(c), they can be related as:

$$C_1(t) = \Theta_{12} C_2(t) + \beta_{12}, \tag{5}$$

where $\Theta_{12}$ is the relative drift between two receivers and $\beta_{12}$ is the offset due to the bias variations. While the clock bias of two receivers are the same, it cancels the offset, implying that $\beta_{12} = 0$.

Following Fig. 4(c), the overall impact of GNSS enabled synchronization on the networking nodes can be practically estimated by an end-to-end timing comparison. Conducting a series of extensive experiments, we will present our results on availability and accuracy of GNSS time synchronization, respectively, in the next two sections.

## VI. AVAILABILITY OF GNSS TIME SOLUTIONS

Availability is an indication of the capability of the system to provide usable service within a specified coverage area. This section aims to understand to what extent the shielding of GNSS signals by physical structures reduces the availability of GNSS position and time solutions. Therefore, we analyze a set of vehicle data collected in some high-rise streets in Brisbane city, Australia, at the rate of 10 Hz. We use consumer-grade receivers to collect the data and versatile GNSS data processing software RTKLIB to process and analyze the collected data. Fig. 5 illustrates the vehicle tracks with GPS, BDS and GPS+BDS services obtained with 4 or more satellites in view, while Fig. 6 plots the number of visible GPS and BDS satellites over the elevation of 10 degrees.

The vehicle tracks in Fig. 5 show that there are not only position outages due to signal blockages, but also wrong position solutions due to the weak geometry. They also indicate that using multiple constellations does improve the availability substantially. Table II summaries that visibility of GNSS satellites in three cases: four or more satellites, 1 to 3, and null (<1). Table III gives the availability percentages of valid GNSS position under different GDOP values. For example, for ⩽6, Table III clearly shows that the availability parameters of valid 4D states (position and time) in high-rise streets are as low as 49.6%, 33.6% and 80.2% with GPS, BDS and GPS+BDS constellations, respectively. In comparison, the visibilities for a minimum of 1 satellite in the three same constellations are 100%, 99.98% and 100%, respectively. This implies that the negative impact of signal blockages on the availability of GNSS time services is much lower than that of GNSS position services. This example shows the high feasibility of GNSS time synchronizations in the high-rise urban areas, particularly with multi-GNSS constellations.

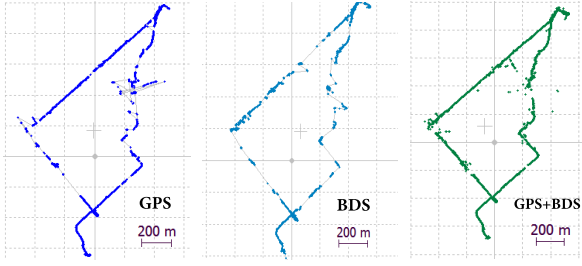

Fig. 5. Vehicle tracks of of GPS, BDS and GPS+BDS on high rising roads.

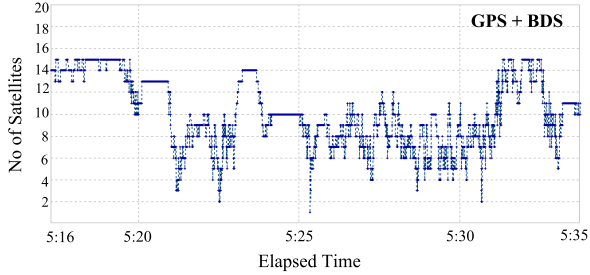

Fig. 6. The number of satellites under the signal coverage of BDS and GPS.

TABLE II
THE NUMBER OF SATELLITES AVAILABLE WITH DIFFERENT
GNSS SERVICE CONSTELLATION

| GNSS System | NSAT ≥ 4 | NSAT = (1 to 3) | NSAT < 1 |
|---|---|---|---|
| GPS | 77.32% | 22.68% | 0% |
| BDS | 82..93% | 17.05% | 0.02% |
| BDS+GPS | 99.25% | 0.75% | 0% |

TABLE III
GDOP WITH DIFFERENT GNSS SERVICES

| GNSS System | NSAT < 4 | GDOP ≤ 6 | GDOP > 6 |
|---|---|---|---|
| GPS | 22.68% | 49.61% | 27.71% |
| BDS | 17.07% | 33.59% | 49.34% |
| BDS+GPS | 0.75% | 80.25% | 19.00% |

## VII. SYNCHRONIZATION ACCURACY OF 1 PPS SIGNALS

This section conducts experiments to demonstrate how accurately consumer-grade GPS receivers are able to synchronize their clocks with 1 PPS signals in VANET.

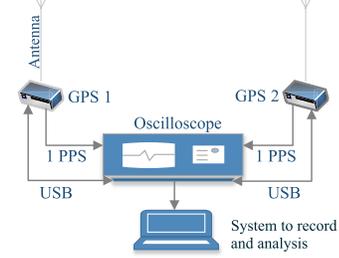

Fig. 7. Experimental setup for synchronization assessment of GPS receivers.

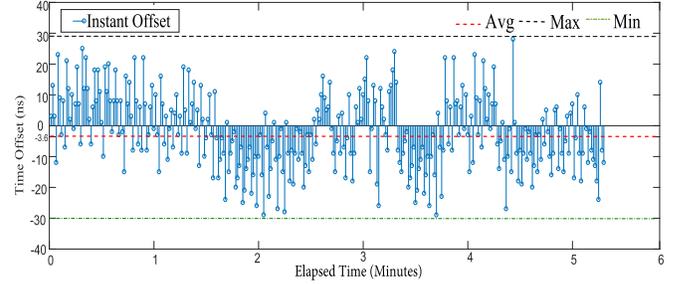

Fig. 8. Time offset distribution of data for 5 min.

### A. Experimental Design

The GPS receivers we have used in our experiments are from U-Blox and Furuno, which are low-cost and consumer-grade. Both U-Blox and Furuno use Temperature Compensated Crystal Oscillator (TCXO) that has excellent stability over a broad range of temperature. They are equipped with a Pluto+ RPT5032A model quartz oscillator. The clock in such a GPS received is designed with an advanced temperature compensation circuit. According to the technical specifications of the product, it offers sub 0.1 ppm frequency stability over an extended temperature range (55°C to 105°C).

Our experimental setup is given in Fig. 7. Two GPS receivers along with two identical antennas are connected to a high resolution oscilloscope, 200MHz Agilent Technology DSO-X-2024A. The oscilloscope measures the offset from the 1 PPS signals of the receivers. A process to calibrate and record experimental data is developed using Lab-View software on a computer system as shown in Fig. 7.

To understand the overall accuracy of different GPS receiver clocks, two scenarios are tested in our experiments. The first scenario considers two GPS receivers of the same model from the same vendor. The other scenario considers two GPS receivers from two different vendors. The 1 PPS pulses from the two receivers are acquired and recorded for further comparative analysis. The experiments are carried out under the temperature (18°C-25°C).

### B. Time Offset Between Receivers of the Same Model

Fig. 8 shows plots of time offset and jitter obtained over 5 minutes between two receivers of the same model from the same vendor. It is seen from this figure that the differences between the 1 PPS outputs of the two receivers change over time randomly. It is observed from this data set that the peak value of |offset+jitter| is 30 ns. The mean value of the offset measurements is calculated as 3.6 ns.

TABLE IV
RELATIVE OFFSET IN ns BETWEEN RECEIVERS OF THE SAME MODEL

| Test Session | Peak | Mean | STD | RMS |
|---|---|---|---|---|
| 5 min | ±30 | 3.6 | 12.67 | 12.67 |
| 1 hr | ±90 | 4.15 | 12.58 | 13.2 |
| 10 hrs | ±40 | 3.8 | 9.73 | 10.4 |
| 24 hrs | ±60 | 1.75 | 12.2 | 12.2 |

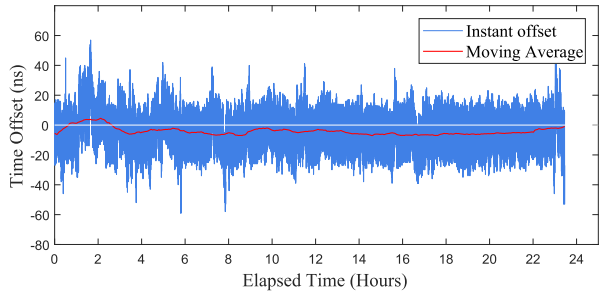

Fig. 9. Time offset between receivers of the same model over a long period.

Table IV summarizes the statistics of peak, mean, standard deviation (STD), and root mean square (RMS) from four independent experimental sessions of 5 min, 1 hour, 10 hours and 24 hours. The results show consistent RMS values between 10 ns and 13 ns. They match the accuracy and precision specifications of the GNSS receiver.

The time offset of PPS solutions between two receivers of the same model over 24 hours is depicted in Fig. 9. As shown in Table IV, in this test case, the STD, mean value and peak value of the the time series are 12.2 ns, 1.75 ns, and ±60 ns, respectively. Again, these results indicate consistent match between the pulse signals from the two receivers.

### C. Time Offset Between Receivers of Different Models

Similar experiments are performed with two GPS receivers of different models from two vendors. Peak, mean, STD and RMS results of the experiments are summarized in Table V summarizes. They are much larger than those in Table IV measured from two GPS receivers of the same model.

The offsets measurements over 24 hours are depicted in Fig. 10. The STD value is measured to be 30 ns. The mean offset is 6.9 ns and peak values are +80 ns and −180 ns. The mean values over a moving data window of 2 hours show offset consistence within ±30 ns.

### D. Further Discussions

It is observed from our experiments that using the same model of GPS receivers in the network enables more accurate time synchronization than using GPS receives of different models. In comparison with the experimental results from GPS receivers of the same model, time synchronization errors almost doubled when GPS receivers from different vendors are used. This is observed from our practical experiments, and is by no means our a claim for general VANET scenarios.

TABLE V
RELATIVE OFFSET IN ns BETWEEN RECEIVERS OF DIFFERENT MODELS

| Test Session | Peak | Mean | STD | RMS |
|---|---|---|---|---|
| 5 min | ±55 | 10.6 | 9.1 | 31 |
| 1 hr | ±45 | 8.6 | 26.4 | 29 |
| 10 hrs | ±60 | 11.8 | 28 | 29.3 |
| 24 hrs | ±180 | 6.9 | 30 | 31.4 |

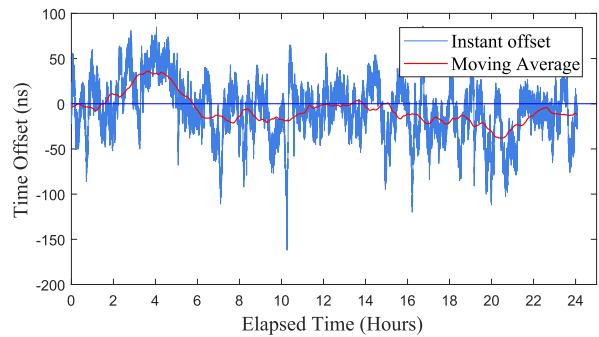

Fig. 10. Time offset between receivers of different models over a long period.

It is inferred that GPS receivers from different vendors may not adopt the same error models and mitigation algorithms in their receiver navigation processors. It is understood that the amplification of relative PVT errors can be minimized if the interoperability requirement for vehicle GNSS receivers is addressed appropriately [53]. Nevertheless, the observed timing errors of tens of nanoseconds can be accommodated for most VANET applications with strict time synchronization accuracy requirements.

Consumer-grade receivers are low-end GNSS devices. Such receivers use C/A code (Single band (L1) Coarse Accusation code) for PVT solutions [49]. Clocks in these receivers with inexpensive quartz oscillators are responsible for receiver clock skews, drifts and noises. In our experiments, we have recorded relative time offsets between two consumer grade receivers. The maximum time offset in our tests is 180 ns, which is the combined effect of individual receiver hardware delays, clock variations, and noises. It is worth mentioning that this relative-receiver time offset is not the timing parameter to be used for further positioning and velocity calculation but a measure of time synchronization capacity between receivers under GNSS services.

The technique of pseudo-range measurement, also known as code-phase, in the Standard Positioning Services (SPS) uses four sophisticated navigational equations and data from four satellites to estimate (4-D estimation) four unknown parameters: three coordinates of the receiver's position and receiver clock bias, i.e., clock error. As long as the receiver gets signal from four or more satellites in view with a standard GDOP value, the uncertainty of clock solutions can be estimated with the pseudo-range errors, $\sigma_P$ and GDOP value, i.e., $\sigma_{(C/A)}$ GDOP. The accuracy of the positioning solutions is similarly given by $\sigma_P$ and GDOP value [54], [55]. The uncertainties of position and clock bias solutions do not affect each other.

As mentioned before, the standard low-cost receiver can also utilize Doppler shift measurements to estimate velocity and clock drift [56]. The design matrix for Doppler and measurements and velocity and clock drift is same as for the pseudo-range measurements and position and clock bias states. Thus the uncertainties of velocity and clock drift parameters do not affect each other but determined by the DOP values and the Doppler measurement errors.

## VIII. Conclusion

GNSS-based time synchronization and its accuracy requirements have not been well understood in VANET. This paper has discussed the significance and requirements of GNSS time synchronization in VANET. After a thorough and careful study, it has identified a number of VANET applications whose operations reply on time synchronization. For these applications, time synchronization has been classified to be either essential or desirable. After reviewing the difficulties of using existing MANET time synchronization techniques in VANET, this paper has presented a feasibility study from theoretic and experimental investigations. Particularly, experimental studies have been conducted to demonstrate the availability and accuracy of GNSS time solutions. Our experimental results have shown that with the advancement of Multi-GNSS constellation, the availability of GNSS time solutions in highly dense high-rise urban areas is as high as 99.98%. The accuracy of GNSS time synchronization has been shown to be at tens of nanoseconds from customer-grade GNSS receivers. It is quite acceptable for most VANET applications with time synchronization requirements. Therefore, GNSS time synchronization is a promising tool in VANET applications.

## References

[1] B. J. Nijssen, "Cooperative intelligent transportation systems building a demonstrator for the CVIS-project on the informatieve Weg," M.S. thesis, Univ. Twente, Enschede, The Netherlands, 2008.

[2] K. Golestan et al., "Vehicular ad-hoc networks (VANETs): Capabilities, challenges in information gathering and data fusion," in Autonomous and Intelligent Systems. Berlin, Germany: Springer, 2012, pp. 34–41.

[3] S. Ghosh, A. Kundu, and D. Jana, "Implementation challenges of time synchronization in vehicular networks," in Proc. IEEE Recent Adv. Intell. Computat. Syst. (RAICS), Sep. 2011, pp. 575–580.

[4] B. Sundararaman, U. Buy, and A. D. Kshemkalyani, "Clock synchronization for wireless sensor networks: A survey," Ad Hoc Netw., vol. 3, no. 3, pp. 281–323, 2005.

[5] I. Skog and P. Handel, "Time synchronization errors in loosely coupled GPS-aided inertial navigation systems," IEEE Trans. Intell. Transp. Syst., vol. 12, no. 4, pp. 1014–1023, Dec. 2011.

[6] K. S. Bilstrup, "Predictable and scalable medium access control for vehicular ad hoc networks," Ph.D. dissertation, Dept. Signals Syst., Chalmers Univ. Technol., Göteborg, Sweden, Tech. Rep. RO16/2009, 2009.

[7] K. S. Bilstrup, E. Uhlemann, and E. G. Ström, "Delay and interference comparison of CSMA and self-organizing TDMA when used in VANETs," in Proc. 7th Int. Conf. Wireless Commun. Mobile Comput. Conf. (IWCMC), 2011, pp. 1488–1493.

[8] H. Cozzetti and R. Scopigno, "Scalability and QoS in MS-Aloha VANETs: Forced slot re-use versus pre-emption," in Proc. 14th Int. IEEE Conf. Intell. Transport. Syst. (ITSC), Oct. 2011, pp. 1759–1766.

[9] S. Wang, A. Pervez, and M. Nekovee, "Converging time synchronization algorithm for highly dynamic vehicular ad hoc networks (VANETs)," in Proc. 7th Int. Symp. Commun. Syst. Netw. Digit. Signal Process. (CSNDSP), Jul. 2010, pp. 443–448.

[10] G. Tian, Y.-C. Tian, and C. Fidge, "Precise relative clock synchronization for distributed control using TSC registers," J. Netw. Comput. Appl., vol. 44, pp. 63–71, Sep. 2014.

[11] G. Karagiannis et al., "Vehicular networking: A survey and tutorial on requirements, architectures, challenges, standards and solutions," IEEE Commun. Surveys Tuts., vol. 13, no. 4, pp. 584–616, 4th Quart., 2011.

[12] Y. L. Morgan, "Notes on DSRC & WAVE standards suite: Its architecture, design, and characteristics," IEEE Commun. Surveys Tuts., vol. 12, no. 4, pp. 504–518, 4th Quart., 2010.

[13] A. Cozzetti, R. Scopigno, and L. Lo Presti, "Tight coupling benefits of GNSS with VANETs," IEEE Aerosp. Electron. Syst. Mag., vol. 26, no. 4, pp. 15–23, Apr. 2011.

[14] K. A. Hafeez, L. Zhao, B. Ma, and J. W. Mark, "Performance analysis and enhancement of the DSRC for VANET's safety applications," IEEE Trans. Veh. Technol., vol. 62, no. 7, pp. 3069–3083, Sep. 2013.

[15] D. Venezuela, C. Liu, L. Wang, and L. Shi, "Improving scalability of vehicle-to-vehicle communication with prediction-based STDMA," in Proc. 80th IEEE Veh. Technol. Conf. (VTC Fall), Sep. 2014, pp. 1–5.

[16] S. Golnarian, J. N. Laneman, and M. D. Lemmon, "On the outage performance of an ieee 802.11 broadcast scheme in vehicular ad hoc networks," in Proc. 54th Annu. Allerton Conf. Commun., Control, Comput. (Allerton), 2016, pp. 101–106.

[17] K. Bilstrup, E. Uhlemann, E. G. Strom, and U. Bilstrup, "On the ability of the 802.11 p MAC method and STDMA to support real-time vehicle-to-vehicle communication," EURASIP J. Wireless Commun. Netw., vol. 2009, pp. 1–13, Jan. 2009.

[18] A. Böhm, "Delay-sensitive wireless communication for cooperative driving applications," Ph.D. dissertation, School Inf. Sci., Comput. Elect. Eng., Halmstad University, Halmstad, Sweden, 2013.

[19] R. Scopigno and H. Cozzetti, "GNSS synchronization in vanets," in Proc. 3rd Int. Conf. New Technol., Mobility Secur. (NTMS), Dec. 2009, pp. 1–5.

[20] H. A. Cozzetti and R. Scopigno, "RR-Aloha+: A slotted and distributed MAC protocol for vehicular communications," in Proc. IEEE Veh. Netw. Conf. (VNC), Oct. 2009, pp. 1–8.

[21] H. Cozzetti, R. Scopigno, L. Casone, and G. Barba, "Comparative analysis of IEEE 802.11 p and MS-Aloha in Vanet scenarios," in Proc. IEEE Asia–Pacific Services Comput. Conf. (APSCC), Dec. 2009, pp. 64–69.

[22] J.-H. Lim, "Understanding STDMA via computer simulation: Feasibility to vehicular safety applications, configurations, and time synchronization errors," EURASIP J. Wireless Commun. Netw., vol. 2016, no. 1, p. 181, 2016.

[23] R. Ben-El-Kezadri and G. Pau, "TimeRemap: Stable and accurate time in vehicular networks," IEEE Commun. Mag., vol. 48, no. 12, pp. 52–57, Dec. 2010.

[24] J. T. Isaac, S. Zeadally, and J. S. Camara, "Security attacks and solutions for vehicular ad hoc networks," IET Commun. J., vol. 4, no. 7, pp. 894–903, 2010.

[25] G. Caporaletti, "The ERSEC project: Enhanced road safety by integrating Egnos-Galileo data with on-board control system," SAE Tech. Paper 2012-01-0285, 2012.

[26] A. Ebner, H. Rohling, M. Lott, and R. Halfmann, "Decentralized slot synchronization in highly dynamic ad hoc networks," in Proc. 5th Int. Symp. Wireless Pers. Multimedia Commun., vol. 2. 2002, pp. 494–498.

[27] Intelligent Transport Systems (ITS): STDMA Recommended Parameters and Settings for Cooperative ITS; Access Layer Part, document ETSI TR 102 861, European Telecommunications Standards Institute, 2012.

[28] J.-Y. Yoon, J.-W. Kim, W.-Y. Lee, and D.-S. Eom, "A TDoA-based localization using precise time-synchronization," in Proc. 14th Int. Conf. Adv. Commun. Technol. (ICACT), 2012, pp. 1266–1271.

[29] K. Golestan, S. Seifzadeh, M. Kamel, F. Karray, and F. Sattar, "Vehicle localization in VANETs using data fusion and V2V communication," in Proc. 2nd ACM Int. Symp. Design (DIVANet), New York, NY, USA, 2012, pp. 123–130.

[30] E.-K. Lee, S. Yang, S. Y. Oh, and M. Gerla, "RF-GPS: RFID assisted localization in VANETs," in Proc. 6th IEEE Int. Conf. Mobile Adhoc Sensor Syst. (MASS), 2009, pp. 621–626.

[31] X. Zhao, V. Ganapathy, N. Pissinou, and K. Makki, "Evaluation of time synchronization over mobile ad hoc networks," in Sensor and Ad Hoc Networks. Boston, MA, USA: Springer, 2008, pp. 221–235.

[32] K. Römer, "Time synchronization in Ad hoc networks," in Proc. 2nd ACM Int. Symp. Mobile Ad Hoc Netw. Comput., 2001, pp. 173–182.

[33] M. Jadliwala, Q. Duan, S. Upadhyaya, and J. Xu, "Towards a theory for securing time synchronization in wireless sensor networks," in Proc. 2nd ACM Conf. Wireless Netw. Secur., 2009, pp. 201–212.

[34] Q. Li and D. Rus, "Global clock synchronization in sensor networks," IEEE Trans. Comput., vol. 55, no. 2, pp. 214–226, Feb. 2006.

[35] S. Ganeriwal, R. Kumar, and M. B. Srivastava, "Timing-sync protocol for sensor networks," in *Proc 1st Int. Conf. Embedded Netw. Sensor Syst.*, New York, NY, USA, 2003, pp. 138–149.

[36] J. Elson, L. Girod, and D. Estrin, "Fine-grained network time synchronization using reference broadcasts," *ACM SIGOPS Oper. Syst. Rev.*, vol. 36, pp. 147–163, Dec. 2002.

[37] S. Ganeriwal, C. Pöpper, S. Čapkun, and M. B. Srivastava, "Secure time synchronization in sensor networks," *ACM Trans. Inf. Syst. Secur.*, vol. 11, no. 4, pp. 23-1–23-35, Jul. 2008.

[38] W. Su and I. F. Akyildiz, "Time-diffusion synchronization protocol for wireless sensor networks," *IEEE/ACM Trans. Netw.*, vol. 13, no. 2, pp. 384–397, Apr. 2005.

[39] P. Sommer and R. Wattenhofer, "Gradient clock synchronization in wireless sensor networks," in *Proc. Int. Conf. Inf. Process. Sensor Netw.*, Washington, DC, USA, 2009, pp. 37–48.

[40] M. Maróti, B. Kusy, G. Simon, and Á. Lédeczi, "The flooding time synchronization protocol," in *Proc. 2nd Int. Conf. Embedded Netw. Sensor Syst.*, New York, NY, USA, 2004, pp. 39–49.

[41] J. van Greunen and J. Rabaey, "Lightweight time synchronization for sensor networks," in *Proc. 2nd ACM Int. Conf. Wireless Sensor Netw. Appl.*, New York, NY, USA, 2003, pp. 11–19.

[42] Q. Xu, T. Mak, J. Ko, and R. Sengupta, "Vehicle-to-vehicle safety messaging in DSRC," in *Proc. 1st ACM Int. Workshop Veh. Ad Hoc Netw.*, 2004, pp. 19–28.

[43] E. Sourour and M. Nakagawa, "Mutual decentralized synchronization for intervehicle communications," *IEEE Trans. Veh. Technol.*, vol. 48, no. 6, pp. 2015–2027, Nov. 1999.

[44] A. Ebner, H. Rohling, R. Halfmann, and M. Lott, "Synchronization in ad hoc networks based on UTRA TDD," in *Proc. 13th IEEE Int. Symp. Pers., Indoor Mobile Radio Commun.*, vol. 4. Sep. 2002, pp. 1650–1654.

[45] W. Chen, J. Sun, L. Zhang, X. Liu, and L. Hong, "An implementation of IEEE 1588 protocol for IEEE 802.11 WLAN," *Wireless Netw.*, vol. 21, no. 6, pp. 2069–2085, Aug. 2015.

[46] P. Loschmidt, R. Exel, and G. Gaderer, "Highly accurate timestamping for ethernet-based clock synchronization," *J. Comput. Netw. Commun.*, vol. 2012, Dec. 2012, Art. no. 152071.

[47] A. Günther and C. Hoene, "Measuring round trip times to determine the distance between WLAN nodes," in *Proc. 4th Int. IFIP-TC6 Netw. Conf.*, Waterloo, ON, Canada, vol. 3462. May 2005, pp. 768–779.

[48] M. A. Lombardi, L. M. Nelson, A. N. Novick, and V. S. Zhang, "Time and frequency measurements using the global positioning system," *Cal Lab, Int. J. Metrol.*, vol. 8, no. 3, pp. 26–33, 2001.

[49] P. Misra and P. Enge, *Global Positioning System: Signals, Measurements and Performance*, 2nd ed. Lincoln, MA, USA: Ganga-Jamuna Press, 2006.

[50] G. Scott and G.-E. Demoz, *GNSS Applications and Methods*, vol. 685. Norwood, MA, USA: Artech House, 2009.s

[51] M. L. Sichitiu and C. Veerarittiphan, "Simple, accurate time synchronization for wireless sensor networks," in *Proc. IEEE Wireless Commun. Netw. Conf. (WCNC)*, vol. 2. Mar. 2003, pp. 1266–1273.

[52] M. Lévesque and D. Tipper, "A survey of clock synchronization over packet-switched networks," *IEEE Commun. Surveys Tuts.*, vol. 18, no. 4, pp. 2926–2947, 4th Quart., 2016.

[53] ARRB-Project-Team, "Vehicle positioning for cooperative-its in Australia (background document)," AustRoads, Sydney, NSW, Australia, Tech. Rep. NT1632, 2013.

[54] R. B. Langley, "Dilution of precision," *GPS World*, vol. 10, no. 5, pp. 52–59, 1999.

[55] C. Specht, M. Mania, M. Skóra, and M. Specht, "Accuracy of the GPS positioning system in the context of increasing the number of satellites in the constellation," *Polish Maritime Res.*, vol. 22, no. 2, pp. 9–14, 2015.

[56] W. Ding and J. Wang, "Precise velocity estimation with a stand-alone GPS receiver," *J. Navigat.*, vol. 64, no. 2, pp. 311–325, 2011.

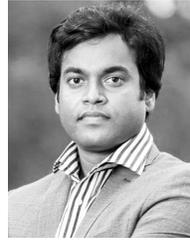

**Khondokar Fida Hasan** (S'13) received the B.Sc. and M.Sc. degrees in electrical and electronics engineering from Islamic University, Bangladesh. He is currently pursuing the Ph.D. degree with the School of Electrical Engineering and Computer Science, Queensland University of Technology, Brisbane, QLD, Australia. His current research interests include wireless communications, network clock synchronization, intelligent transportation systems, vehicular ad-hoc networks, and global navigation satellite systems.

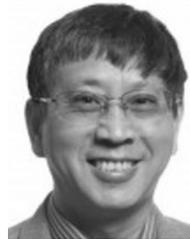

**Yanming Feng** received the Ph.D. degree in satellite geodesy from the Wuhan Technical University of Surveying and Mapping (currently part of Wuhan University), Wuhan, China. He is currently a Professor at the School of Electrical Engineering and Computer Science, Queensland University of Technology, Brisbane, QLD, Australia. His research interests include wide-area global navigation satellite systems positioning, multiple-carrier ambiguity resolution, and real-time kinematic positioning and applications.

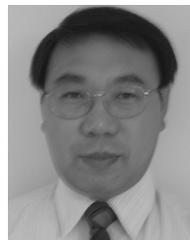

**Yu-Chu Tian** (M'00) received the Ph.D. degree in computer and software engineering from the University of Sydney, Sydney NSW, Australia, in 2009, and the Ph.D. degree in industrial automation from Zhejiang University, Hangzhou, China, in 1993. He is currently a Professor at the School of Electrical Engineering and Computer Science, Queensland University of Technology, Brisbane, QLD, Australia. His research interests include big data computing, distributed computing, cloud computing, real-time computing, and computer networks.